# Increasing the coherence time of single electron spins in diamond by high temperature annealing

Boris Naydenov[1], Friedemann Reinhard[1], Anke Lämmle[1], V. Richter[2], Rafi Kalish[2], Ulrika F. S. D'Haenens-Johansson[3], Mark Newton[3], Fedor Jelezko[1] and Jörg Wrachtrup[1]

1) 3 Physikalisches Institut and Research Center SCoPE, University of Stuttgart, Stuttgart 70659, Germany

2) Solid State Institute, Technion City, Haifa Israel

3) Department of Physics, University of Warwick, Coventry CV4 7AL, England

Negatively charged Nitrogen-Vacancy (NV⁻) centers in diamond produced by ion implantation often show properties different from NVs created during the crystal growth. We observe that NVs created from nitrogen ion implantation at 30 - 300 keV show much shorter electron spin coherence time $T_2$ as compared to the "natural" NVs and about 20 % of them show switching from NV⁻ to NV⁰ . We show that annealing the diamond at T = 1200 °C substantially increases $T_2$ and at the same time the fraction of NVs converting from NV⁻ to NV⁰ is greatly reduced.

In the recent years negatively charged Nitrogen Vacancy (NV⁻) defect centers in diamond have attracted the attention of many research groups due to their unique properties (for simplicity we denote below the negatively charged NV⁻ center as NV). The NV consists of an interstitial nitrogen and vacancy next to it and an extra electron attached to this complex. It thus has a spin 1 triplet ground state. The fact that the electron spin of a single NV defect can be detected and manipulated even at room temperature (RT)[1,2] and that it has extremely long coherence time ($T_2$ = 1.8 ms[3] at RT) makes them ideal solid state quantum bits (qubits) for a solid state quantum computer (QC). Another promising application of the NVs is to use them as sensors for measuring ultra low magnetic fields with nanometer precision[4,5]. The first experiments were performed on NVs created during growth of the diamond crystal, but these defects were later produced by ion implantation offering better control on their formation efficiency and on their location[6]. In this method first nitrogen ions are implanted into pure diamond and the vacancies are created during their slowing down process, followed by annealing (T > 800 °C in vacuum (or in inert gas like $N_2$ or Ar) for several hours. During this process the vacancies start to diffuse (see[9] for more details about their diffusion) and form an NV center with the implanted interstitial nitrogen atom. In order to distinguish between "natural" and implanted NVs, ¹⁵N (natural abundance 0.4 %, nuclear spin I = 1/2) can be implanted instead of ¹⁴N (99.6 %, I = 1). Different nuclear spins induce different splitting through the hyperfine coupling in the Optically Detected Magnetic Resonance (ODMR) spectrum, thus making it possible to determine the source of the NV[7]. It is important to create these color centers on demand, since for a scalable QC the positioning of the qubits is crucial. The first quantum register based on two implanted single NVs has been already demonstrated[8]. Surprisingly, the coherence time $T_2$ of the pair was much shorter ($T_2^A$ = 110 μs and $T_2^B$ = 2 μs) as compared to $T_2$ of the NV produced during the crystal growth (for a ¹²C enriched diamond, $T_2$ approximates 1 ms) which did not allow to entangle the two qubits. This short $T_2$ could be explained by the presence of defects around the NV center, which are created during the penetration of the nitrogen ion through the diamond crystal lattice and are not destroyed during the annealing (see[10] for a review). These defects couple to the electron spin of the NV center, thus inducing decoherence. Furthermore, the negatively charged NV center has the tendency to lose its extra electron, thus converting into an $NV^0$ a process also possibly driven by the presence of defects in its surrounding. Hence a new method has to be developed in order to reduce the concentration of various defects around the NV center, thus enhancing $T_2$ and reducing the discharging effects.

In this Letter we demonstrate the increase of the coherence time and increasing the fraction of $NV^-$ of implanted single NV centers by annealing the diamond at high temperature (T = 1200 °C) under nitrogen gas. First the NVs were produced by implanting $^{15}N^+$, at an angle of 7 degrees away from the perpendicular to the surface, to avoid implanted ion channeling, with four different energies - 30, 60, 150 and 300 keV with dose $1 \times 10^8$ ions/cm$^2$, followed by annealing in vacuum at T = 800 °C for three hours. Afterwards the sample was kept for 3 hours in a boiling acid mixture consisting of sulphuric, nitric and perchloric acid (1:1:1) in order to clean the surface from any graphite residues or other impurities formed during the annealing. The ODMR measurements were performed on a home built confocal microscope. The electron spin coherence time $T_2$ was measured by using the standard ODMR Hahn echo pulse sequence[1]. The number of photon emitters in the confocal spots was calibrated by recording the second order correlation function of the fluorescence of single NV centers.

The production efficiency of NV centers (ratio of the implanted nitrogen ions to the observed NVs) in this energy range has not been studied previously and it was expected to lie between 10-40 %[14]. Indeed we observe that the yield increases from 26 % to 37 % with increasing N ion implantation energy (see table 1). This result was expected since it has been shown earlier that the main limitation of the NV creation yield is the low vacancy concentration around the implanted nitrogen ion[14,15]. Indeed, by loading the volume around the N implants by vacancies induced by C ion implantations we could show that the NV formation efficiency has increased[15]. By increasing the energy, the number of vacancies produced by each ion also increases, resulting in higher yield.

| Energy, keV | Depth, nm | Straggle, nm | NV Yield, % |
| --- | --- | --- | --- |
| 30 | 40 | 11 | 26 |
| 60 | 74 | 17 | 28 |
| 150 | 166 | 27 | 30.8 |
| 300 | 301 | 37 | 37 |

Table1. Production efficiency of NV as a function if the implantation energy. The position and straggle (≈ half width at half maximum) of the nitrogen ions in the diamond have been calculation with SRIM 2010[11]

Another goal of this study was to investigate how the coherence time $T_2$ depends on the implantation energy and accordingly on the distance from the NV center to the diamond surface. To our surprise we observed that $T_2$ was not significantly different for all four energies. Moreover, the $T_2$ values can be divided in two distinct groups. In the first one we can sort out about 40 % of the NVs, which show short $T_2$ (below 10 μs). For a second part (about 40 %) we measured $T_2$ longer than 50 μs. Echo decays typical for these two groups are depicted in Fig. 1, where the implantation energy was 300 keV. This data is summarized in Fig. 2

For the last group of NVs (about 20 %) we could not detect any Hahn echo decay (although we could record an ODMR spectrum), but we observed that the fluorescence of the NV steadily decays below the baseline on a time scale around 10 μs. By detecting the fluorescence from NV (Zero Phonon Line (ZPL) 638 nm) and NV$^0$ (ZPL 575 nm) in two separate channels, we could see that while the signal from NV$^-$ decreases, the fluorescence from NV$^0$ increases proportionally. This result is explained by discharging the NV$^-$ to NV$^0$ in the absence of optical excitation. If the laser pulse is constantly exciting the NV, than the reverse process is observed and NV recharges. This phenomenon has been already reported by Gaebel et al.[16] for single NVs created by implantation of nitrogen at E = 7 keV. The fact that NV$^-$ is discharging would suggest that there are some electron acceptors in the surrounding which readily take the electron from the NV after its excitation with the laser pulse. The statistical data for E = 300 keV (overall 34 centers) reveals that 18 % of the NVs show discharging.

One way to remove the various paramagnetic defects created around the NV center during the implantation would be to anneal the diamond at higher temperature (T > 1000 °C) in vacuum or in an inert gas. Many of these defects, (for example O1, R4, R5, R9, R10[10]) are expected to be destroyed under these conditions. In order to test this idea we annealed the sample at T = 1200 °C for 12 hours. Nitrogen gas was flown during the annealing in order to suppress converting the diamond surface into graphite. Additionally the diamond was placed in an alumina holder enveloped in diamond powder (Element Six, Micron+ MDA M12) which was supposed to act as sacrificial layer. However, after acid cleaning we observed that the number of NVs in the areas of shallow implantations (implanted with 30 and 60 keV ions) has been greatly reduced. This means that despite our efforts not to damage the surface, it has graphitized or it has been etched away down to a depth of about 70 nm. Nevertheless, Hahn echo measurements of the centers in the 150 keV and 300 keV regions revealed that the coherence time has been significantly improved. The same three types of NVs differing in lifetime can be again identified, although the fraction of the NVs showing short $T_2$ as well the NVs showing discharging have been both reduced by a factor of three - from 18 % down to 6 %. At the same time we observe about two-fold increase in the NVs with long $T_2$. These results are depicted in Fig. 2.

A possible defect which could cause decoherence of the NV is the nearest neighbor divacancy R4 (or W6)[17]. This paramagnetic center (electron spin S = 1) is easily formed by ion implantation in diamond, followed by annealing. It can be destroyed if the sample is kept longer time at temperature above T = 1100 °C[10] . There are also other paramagnetic defects (like R6, R7, R8, R12, W16, W17, W18[10] ) in diamond that remain stable even at temperatures above 1200 °C which can still limit the coherence time of the NV. Annealing at even higher temperatures would probably remove the remaining defect centers.

In summary, we have shown that $T_2$ of NVs created by nitrogen ion implantation does not depend on the ion energy in the range 30 - 300 keV. A large part of the centers have short $T_2$ , whereas a smaller fraction has long coherence time. Some of the NVs change their charge state from negative to zero. After annealing the sample at T = 1200 °C we observe that the fractions of NVs showing short $T_2$ and discharging are greatly reduced. We believe that this method of high temperature annealing for improving the properties of the implanted NVs can be useful for the application of the NV as ultra low field nanoscale magnetometer as well as a qubit.

We acknowledge the financial support from the German-Israeli Foundation, Landesstiftung Baden-Württemberg and VolkswagenStiftung.

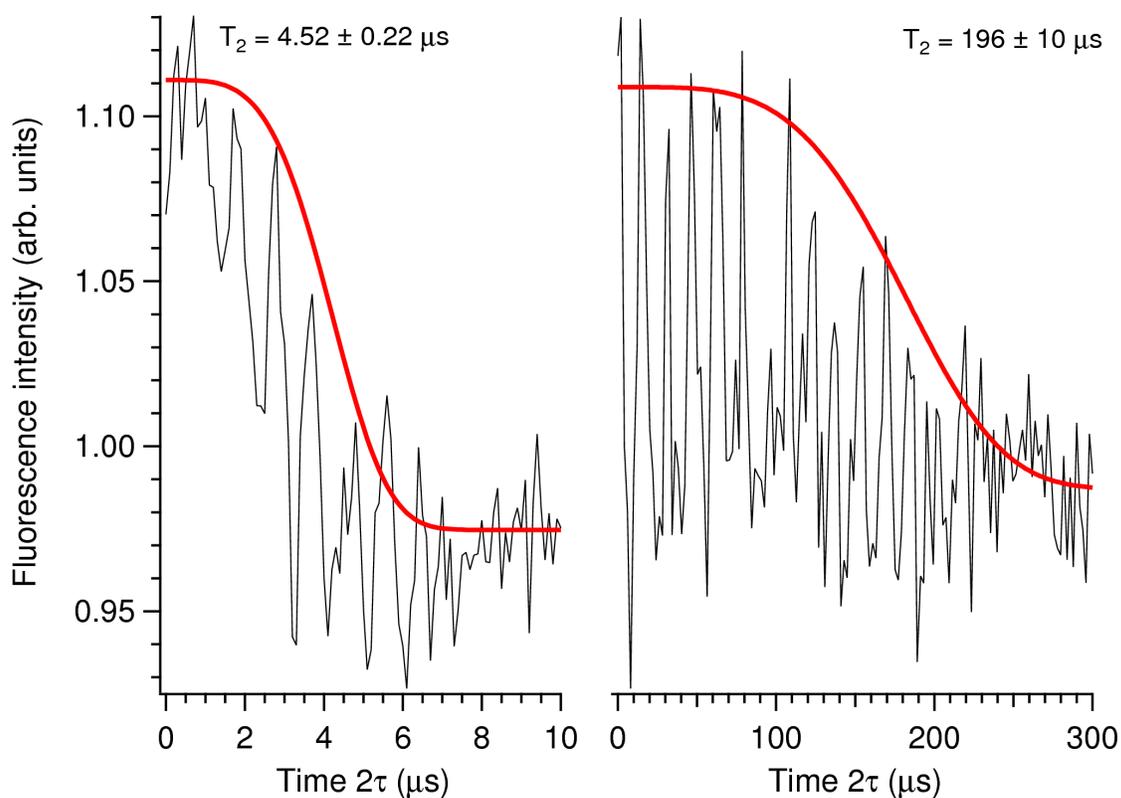

Figure 1: Hahn echo decays of single NV centers produced by implanting nitrogen ions with E = 300 keV showing two types of $T_2$ - short (left) and long (right). The oscillations are induced by coupling to the $^{15}N$ nitrogen nucleus (left) and $^{13}C$ spin bath (right). See[12,13] for more details.

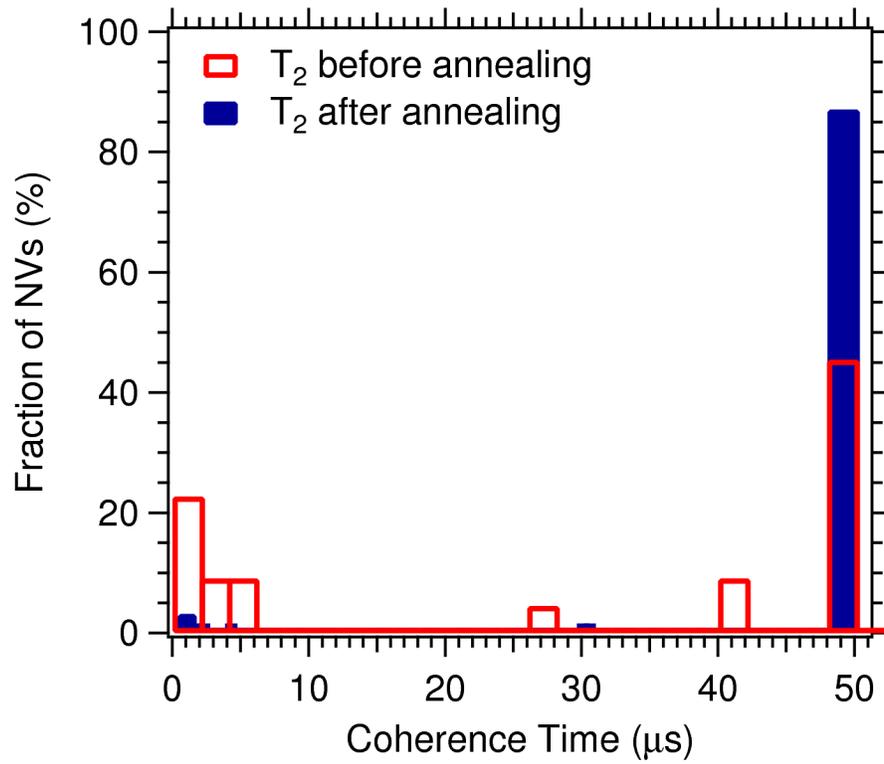

*Figure 2: Electron spin coherence time $T_2$ of NV centers before (red) and after annealing (blue). Note that 50 μs is the lowest limit for the NVs with long $T_2$.*